\documentclass{article} 

\usepackage{graphicx} 
\usepackage{mathrsfs}
\usepackage{epsfig}
\usepackage{amssymb}
\usepackage{amsmath}

\newcommand{\D}{\mathrm{d}}
\newcommand{\DD}{\mathrm{D}}
\newcommand{\E}{\mbox{1\hspace{-3 pt}I}}

\newcommand{\R}{\mathbb{R}}
\newcommand{\Exp}[1]{\text{e}^{#1}}
\newcommand{\binomial}[2]{\left(\begin{array}{c} #1 \\ #2 \end{array} \right)}
\newcommand{\pd}{\partial}
\numberwithin{equation}{section}

\begin{document}
\title{Relativistic Fluctuation Theorems}

\author{Axel Fingerle, \\
axel.fingerle@ds.mpg.de, \\
Max Planck Institute for Dynamics and Self-Organization \\
Bunsenstr. 10, Germany - 37073 G\"ottingen}

\date{\today}

\begin{abstract}
To reveal how nonequilibrium physics and relativity theory
intertwine, this articles studies relativistic Brownian motion
under cosmic expansion. Two fluctuation theorems for the entropy
$\Delta s$, which is locally produced in this extreme
nonequilibrium situation, are presented and proven. The first,
$\left<\Exp{-\Delta s}\right>=1$, is a generalization of the
second law of thermodynamics, that remains valid at relativistic
particle energies and under high cosmic expansion rates. From this
relation follows, that the probability to observe a local
reduction of entropy is exponentially small even if the universe
was to recollapse. For the special case of the Einstein-de~Sitter
universe an additional relation,  $\left<\Exp{-\Delta s-\Delta
h}\right>=1$, is derived which holds simultaneously with the first
relation and where $\Delta h$ is proportional to the Hubble
constant. Furthermore, the fluctuation theorems are shown to
provide a physical criterion to resolve the known discretization
dilemma arising in special-relativistic Brownian motion. Explicit
examples and a general method for the computation of non-Gaussian
entropy fluctuations are provided.
\end{abstract}

\maketitle

\section{Introduction} The physical basis of the direction
of time has been discussed since Boltzmann's H-Theorem in 1872. A
priori, the thermodynamic arrow of time has to be distinguished
from the possibility of a prime direction of time defined by the
expansion of the universe \cite{Zeh}. By now we know that due to
the dominating dark energy component of about $72\%$, our universe
is very likely to expand forever \cite{WMAP3yr}. Yet, the
fascinating cosmological arrow of time could not be based on firm
theoretical ground \cite{Hawking1, Page, Hawking2, Spec1, Spec2}.
So one may still ask: Is it a mere coincidence that our memory
strictly refers to times when the size of the universe was
smaller? Put in physical terms, the guiding question of this
article is: Does the cosmic expansion rate effect the production
of entropy by nonequilibrium processes? While a general theory of
nonequilibrium thermodynamics does not exist, fluctuation theorems
(FTs) provide a unique starting point to develop the means to
address such a fundamental question. First progress in the
description of entropy production and giant fluctuations beyond
linear-response was made in the 1970s \cite{BochkovKuzovlev}, and
major advance was achieved in recent years with the derivation of
FTs for various classes of systems \cite{Evans, GallavottiCohen,
Kurchan, Zon, Udo, Jarzynski}. FTs generalize the second law of
thermodynamics. The second law states that the Gibbs entropy $S$
of an ensemble may not decrease,
\begin{eqnarray}
  \Delta S \ge 0 \text{ at any time.} \label{SL}
\end{eqnarray}
The FTs naturally extend the concept of entropy and allow
statements about the probability to observe isolated
``violations'' of (\ref{SL}). To this end the FTs assign a change
$\Delta s$ of entropy to an observation of few or even single
particles. When a nonequilibrium system of finite size is
observed, the entropy $\Delta s$ produced within a certain time
interval is a fluctuating quantity. The founders of statistical
mechanics, L.~Boltzmann and J.W.~Gibbs were well aware that the
second law holds only for the entropy $\Delta S=\left< \Delta s
\right>$ of an infinite ensemble. The angle brackets denote the
ensemble average over observations of equal systems. Boltzmann
mentioned the possibility of ``violations'', $\Delta s < 0$, in
his famous reply to the Poincar\'e recurrence objection (in a written argument with E.~Zermelo) and
designated the second law as a theorem of probability
(``Wahrscheinlichkeitssatz''), emphasizing that the second law
cannot be expected to hold for few particles \cite{Boltzmann}.
Loco citato, he referred to Gibbs \cite{Gibbs} who had concluded:
``The impossibility of an incompensated decrease of entropy seems
to be reduced to an improbability.'' It is this improbability that
is quantified by FTs.

For the steady state of strongly chaotic systems the detailed FT,
\begin{eqnarray}
  \frac{\text{Prob}(\Delta s=+a \ {k_{\text{B}}})}
       {\text{Prob}(\Delta s=-a \ {k_{\text{B}}})} =
  \Exp{
  {a}
  }
  \text{ for any $a$}, \label{FT1}
\end{eqnarray}
was proven in the limit of infinite observation time
\cite{GallavottiCohen}. The detailed FT (\ref{FT1}) was also
derived in \cite{Kurchan} for a non-relativistic particle in
contact with a heat bath at temperature $T$. Initially, only the
external change in the bath entropy, $\Delta
s_{\text{e}}=\nolinebreak\Delta Q / T$ with the energy $\Delta Q$
dissipated into the surrounding bath, was taken into account
\cite{Kurchan, Zon}. In \cite{Udo} it was pointed out that when
the particle is assigned an intrinsic entropy
$s_{\text{s}}=\nolinebreak- k_{\text{B}} \ \ln P$ (with the
particle's phase space density $P$), the sum of intrinsic and
external entropy, $\Delta s=\Delta s_{\text{s}}+\Delta
s_{\text{e}}$, obeys the FT (\ref{FT1}) even for finite
observation time\footnote{In the earlier FT of Evans and Searles a
similar term was added to the dissipation function (Eq. (2.6) in
\cite{EvansSearles}), that is not present in the Gallavotti-Cohen
FT \cite{GallavottiCohen}, resulting in a FT for chaotic systems
which holds for finite time. For the stochastic formulation of
finite time FTs such a term was considered in \cite{Maes2},
Eqs.~(5.9) and (5.11).}. This is the definition of entropy applied
throughout the article. Furthermore, for non-stationary states in
the presence of time-dependent driving forces, an integral FT of
the form
\begin{eqnarray}
  \left< \Exp{-\Delta s/k_{\text{B}}} \right> = 1, \label{FT2}
\end{eqnarray}
was proven and linked to the Jarzynski relation \cite{Jarzynski}.
Technically, the averaging over observations,
$\left<\dots\right>$, is a path integral over trajectories,
which is explained in the following section.

The detailed FT (\ref{FT1}), the integral FT (\ref{FT2}), and the
second law of thermodynamics (\ref{SL}) form a consistent hierarchy of statements:
from Eq.~(\ref{FT1}) follows (\ref{FT2}) by integrating over $a$,
and Eq.~(\ref{FT2}) implies (\ref{SL})
by virtue of the Jensen inequality.

We proceed as follows. In Sec.~\ref{Herleitung}, FTs for general
processes are derived. The recent unification \cite{Debbasch1,
DunkelHaenggi1, DunkelHaenggi2} of Einstein's 1905 publications on
Brownian motion \cite{Einstein1} and special relativity
\cite{Einstein2} is briefly reviewed in Sec.~\ref{RBM}. Based on
these findings, the results of Sec.~\ref{SRResults} are twofold.
First, for the relativistic Brownian processes of
\cite{DunkelHaenggi1} and \cite{DunkelHaenggi2}, we reconcile the
FTs (\ref{FT1}) and (\ref{FT2}), which have become a paradigm of
nonequilibrium physics, with special relativity. For the similar
process suggested in \cite{Debbasch1}, FTs follow by analogous
reasoning. In \cite{DunkelHaenggi1} and \cite{DunkelHaenggi2} it
was pointed out that the relativistic time dilation leads to
multiplicative coupling, necessitating a careful choice of the
discretization rule. We show explicitly that there is one
relativistic detailed FT (\ref{FT1}) and one relativistic integral
FT (\ref{FT2}) valid for all choices. Second, we shall find the
physically correct expression for the entropy production following
from relativistic FTs when the H\"anggi-Klimontovich
discretization rule is applied. In Sec.~\ref{GRResults} we go
beyond special relativity with a set of two general-relativistic
integral FTs for the cosmological standard model. These expose
clearly the role of cosmic expansion in entropy production. We
shall identify the entropy production which is solely due to the
Hubble expansion of space. Such entropy producing processes
dominate when the expansion rate of the universe exceeds the
particle scattering rate, for instance in an early inflationary
phase after the big bang. The Secs.~\ref{SRResults} and
\ref{GRResults} conclude each with examples where we explicitly
compute the non-Gaussian fluctuations $P(\Delta s)$, $P(\Delta
s_{\text{s}})$ and $P(\Delta s_{\text{e}})$ of entropy production.

\section{Stochastic formulation of the fluctuation theorem}
\label{Herleitung} This section gives a general derivation of the
FT for stochastic processes and emphasizes that every broken
symmetry implies a FT.

\subsection{The integral fluctuation theorem}
Let ${\bf \Gamma}(t)$ denote the state of the observed system,
which performs a time continuous stochastic process under the
influence of a thermal environment. Its stochastic dynamics are
described completely by the probability distribution $P[{\bf
\Gamma},{\bf C}]$, which gives the probability to observe a
certain system trajectory ${\bf \Gamma}$. This probability depends
on the environmental conditions: $\bf C$ describes a set of
external parameters, such as the environmental temperature $T(t)$,
external forces (for example acting on a charged system by an
electric field ${\bf E}(t)$), or -- as we shall consider finally
in the general-relativistic case -- the curvature of spacetime.
All these external parameters may vary during the process, so that
${\bf C}(t)$ is a deterministic protocol. The general idea
underlying stochastic formulations of fluctuation theorems is as
follows. Consider an arbitrary transformation $\cal T$, which does
not leave the physical dynamics (represented by a Langevin or
Fokker-Planck equation) invariant. While $P[{\bf \Gamma},{\bf C}]$
describes the dynamics of the original stochastic system, the
transformed stochastic dynamics will be given by another
probability distribution $\tilde P[{\bf \Gamma},{\bf C}]$.
Assuming $P$ and $\tilde P$ to have the same support, we define
\begin{eqnarray}
  \Delta s[{\bf \Gamma},{\bf C}] \equiv k_{\text{B}} \ln P[{\bf \Gamma},{\bf C}] - k_{\text{B}} \ln \tilde P[{\bf \Gamma},{\bf C}] \ , \label{sDef}
\end{eqnarray}
to quantify the symmetry breaking of the transformation $\cal T$
for every trajectory ${\bf \Gamma}$. For the quantity $\Delta s$
defined in (\ref{sDef}), an integral FT of the form (\ref{FT2}) is a mathematical identity,
\begin{equation*}
\left< \Exp{-\Delta s/k_{\text{B}}} \right> = \int {\cal D} \left[ {\bf \Gamma} \right] \ P\left[{\bf \Gamma},{\bf C} \right]
  \ \Exp{-\Delta s/k_{\text{B}}}
                                 = \int {\cal D} \left[ {\bf \Gamma} \right] \ \tilde P\left[{\bf \Gamma},{\bf C} \right] =
                                 1\ .
\end{equation*}
The path integration $\int {\cal D}\left[ {\bf \Gamma} \right]$ covers all continuous functions $\bf \Gamma$, weighted by the probability $P[{\bf \Gamma}, {\bf C}]$.

We are interested in a fluctuation theorem that quantifies the
irreversibility of the stochastic process. This is why we choose the transformation
$\cal T$ to be time reversal\footnote{If ${\bf \Gamma}$ is the
phase space vector $({\bf x},{\bf p})$, the momenta are inverted,
$\tilde {\bf \Gamma}(t)=({\bf x(-t)},-{\bf p}(-t))$.},
\begin{eqnarray*}
  && \tilde P[{\bf \Gamma},{\bf C}] = P[\tilde {\bf \Gamma},\tilde {\bf C}], \\
  && \text{with} \ \tilde {\bf \Gamma}(+t)={\bf \Gamma}(-t) \\
  && \text{and}  \ \tilde {\bf C}(+t)={\bf C}(-t) \text{ for all } t.
\end{eqnarray*}
To get a result on the total entropy production, the
transformation $\cal T$ acts globally by reversing both, the
stochastic system trajectory $\bf \Gamma$ and the time dependence of
the environment $\bf C$.

The probability $P[{\bf \Gamma},{\bf C}]$ to observe a stochastic
trajectory $[{\bf \Gamma}]_{-\tau}^{+\tau}$ in  the time
interval $(-\tau,+\tau)$ depends on the initial conditions, which are given by ${\bf
\Gamma}(-\tau)$ for a Markov process or by the history $[{\bf
\Gamma}]_{-\tau-T}^{-\tau}$ for a system with memory time $T$
(which may be infinite):
\begin{eqnarray}
  &&P[{\bf \Gamma},{\bf C}]=P_{\text{in}} \  P_{\text{F}} \  , \label{Pforw} \\
  &&\nonumber \text{with }
  P_{\text{in}}=\left\{\begin{array}{ll}
                              P{{\left({\bf \Gamma},{\bf C}\right)}\vert}_{-\tau}, &\text{ if Markovian} \\
                              P[{\bf \Gamma},{\bf C}]_{-\tau-T}^{-\tau}, &\text{ if with memory}
                       \end{array} \right. \\
  &&\nonumber \text{and }
  P_{\text{F}}=P\left([{\bf \Gamma},{\bf C}]_{-\tau}^{+\tau} \ \vline \
  \text{in} \right) \ .
\end{eqnarray}
We refer to the initial state
or the history
as the \mbox{in-state} of the system, which is distributed
according to the first factor $P_{\text{in}}$ in (\ref{Pforw}).
The second factor $P_{\text{F}}$ is the (forward) propagator on the time
interval $(-\tau,+\tau)$ under the influence of the thermal
environment. Analogously, the time-reversed probability is written
as
\begin{eqnarray}
\tilde P[{\bf \Gamma},{\bf C}]=\tilde P[\tilde {\bf \Gamma},\tilde {\bf C}]=P_{\text{out}} \  P_{\text{R}} \ . \label{Preve}
\end{eqnarray}
Inserting (\ref{Pforw}) and (\ref{Preve}) in (\ref{sDef}), $\Delta
s$ decomposes into the sum
\begin{eqnarray}
  && \Delta s = \Delta s_{\text{s}} + \Delta s_{\text{e}}\  , \label{decompTot} \\
  && \text{with }
  \Delta s_{\text{s}}=-k_{\text{B}}\ln P_{\text{out}}+k_{\text{B}}\ln P_{\text{in}}, \label{decompSys} \\
  && \text{and }
  \Delta s_{\text{e}}=k_{\text{B}} \ln \frac{P_{\text{F}}}{P_{\text{R}}} \ . \label{decompEnv}
\end{eqnarray}
The first term (\ref{decompSys}) is the change of the system
entropy $s_{\text{s}}=-k_{\text{B}} \ln P_{{j}}$ as the system
state changes from ``$j=\text{in}$'' to ``$j=\text{out}$''. The
expression $s_{\text{s}}(P_{{j}})=-k_{\text{B}} \ln P_{{j}}$ for
the system entropy was suggested in \cite{Udo} for a Markov
process and is a widely accepted definition because $s_{\text{s}}$
resembles the Boltzmann entropy and
$S_{\text{s}}=\left<s_{\text{s}}\right>=-k_{\text{B}} \left< \ln
P_{{j}} \right>$ coincides with the Gibbs entropy of the ensemble $P_j$.

The second term (\ref{decompEnv}) is a Crooks relation
\cite{Crooks} defined by the forward and reversed time evolution
under the stochastic influence of the thermal environment. We have
to show that $\Delta s_{\text{e}}$ as introduced in
(\ref{decompEnv}) equals exactly the entropy produced in the
thermal environment, so that $\Delta s=\Delta s_{\text{s}}+\Delta
s_{\text{e}}$ is the total entropy production. It is the objective
of this article to evaluate (\ref{decompEnv}) for a thermal
environment at relativistic energies to adjudicate on
the physical interpretation as environmental entropy.

For Markov processes, such as the relativistic Brownian motion
discussed in the next section, the forward and reverse propagators
are infinite products of transition probabilities,
\begin{eqnarray*}
P_{\text{F}} &=& \lim_{n\rightarrow \infty} \prod_{k=1}^n
P_{\text{trans}}^{\Delta
t_k}\left({\bf \Gamma}_{k-1} \mapsto {\bf \Gamma}_k, {\bf C}(\Delta t_k) \right) \text{ and} \\
P_{\text{R}} &=& \lim_{n\rightarrow \infty} \prod_{k=1}^n
P_{\text{trans}}^{\Delta t_k}\left({\bf \Gamma}_{k} \mapsto {\bf
\Gamma}_{k-1}, {\bf C}(\Delta t_k) \right) \ ,
\end{eqnarray*}
so that entropy production is local in time: For the environmental entropy follows
\begin{eqnarray}
&& \Delta s_{\text{e}} = \int_{-\tau}^{+\tau} {\dot s}_{\text{e}}(t) \ \D t \quad \text{with} \nonumber \\
&& {\dot s}_{\text{e}}(t) \ \D t = k_{\text{B}} \ln
\frac{P_{\text{trans}}^{\D t}\left({\bf \Gamma}^- \mapsto {\bf
\Gamma}^+ ,{\bf C}(t)\right)}{P_{\text{trans}}^{\D t}(\tilde{\bf
\Gamma}^- \mapsto \tilde{\bf \Gamma}^+,{\bf C}(t))} \ , \qquad
\label{decompEnvMark}
\end{eqnarray}
and the change in system entropy is
$ \Delta s_{\text{s}} = s_{\text{s}}(+\tau)-s_{\text{s}}(-\tau)
$ {with} $ s_{\text{s}}(t) = -k_{\text{B}} \ \ln P({\bf \Gamma}(t),{\bf C}(t))$.
The probability density $P({\bf \Gamma},t)=P({\bf \Gamma},{\bf C}(t))$ evolves according to the
continuity equation,
\begin{eqnarray}
\pd_t P({\bf \Gamma},t) + \nabla_{\bf \Gamma} \circ {\bf j}({\bf
\Gamma},t) = 0 \ , \label{ContiEq}
\end{eqnarray}
with the probability current
\begin{eqnarray}
{\bf j}({\bf \Gamma},t)=\sum_{n=0}^\infty \frac{\left(-\nabla_{\bf
\Gamma} \right)^n}{n!} \circ {\bf M}_{n+1} ({\bf \Gamma},t) \
P({\bf \Gamma},t) \ . \label{Strom}
\end{eqnarray}
The Helfand moments ${\bf M}_n$ are tensors of order $n$ which are related
to the transition probability $P_{\text{trans}}^{\D t}$ by \cite{Risken}
\begin{equation*}
 {\bf M}_n({\bf \Gamma},t) \ \D t  = \int \left({\bf
\Gamma}'-{\bf \Gamma}\right)^n \ P_{\text{trans}}^{\D t}\left({\bf
\Gamma}\mapsto {\bf \Gamma}',{\bf C}(t)\right) \ \D {\bf \Gamma}'
\ .
\end{equation*}
The higher moments are present only if the heat bath in which the
system is embedded is out of equilibrium. A possible system for
relativistic Brownian motion is an electron which couples by
Compton scattering to a gas of photons. We assume that such a heat
bath is in local equilibrium so that we have a well-defined
temperature $T({\bf x},t)$ yielding an isotropic diffusion ${\bf
M}_{2} \propto T \E$ with vanishing higher moments, ${\bf M}_n=0$
for $n>2$. Equation (\ref{ContiEq}) then reduces to the
Fokker-Planck equation and the transition probabilities are
Gaussian.

\subsection{The detailed fluctuation theorem} \label{GeneralDFT}
While the integral FT derived above holds for arbitrary
environmental conditions $\bf C$, the stronger detailed FT
(\ref{FT1}) holds if the deterministic protocol is invariant under
time-reversal, ${\bf C} = \tilde {\bf C}$. The general derivation
for the quantity $\Delta s$ defined in (\ref{sDef}) is also done
conveniently by path integration. The probability to observe a
production of entropy $\Delta s=a \ k_{\text{B}}$ is
\begin{eqnarray*}
                         && \text {Prob}(\Delta s=a \ k_{\text{B}}) \\
                         &=& \int P[{\bf \Gamma},{\bf C}] \ \delta (\Delta s[{\bf \Gamma},{\bf C}]=a \ k_{\text{B}}) \ {\cal D}[{\bf \Gamma}] \\
                         &=& \int P[\tilde {\bf \Gamma},\tilde {\bf C}] \ \Exp{\Delta s /k_{\text{B}}}
                                                          \ \delta (\Delta s[{\bf \Gamma},{\bf C}]=a \ k_{\text{B}}) \ {\cal D}[{\bf \Gamma}] \\
                         &=& \Exp{a} \ \int P[\tilde {\bf \Gamma},\tilde {\bf C}]
                                                          \ \delta (\Delta s[{\bf \Gamma},{\bf C}]=a \ k_{\text{B}}) \ {\cal D}[{\bf \Gamma}] \\
                         &=& \Exp{a} \ \int P[\tilde {\bf \Gamma},\tilde {\bf C}]
                                                          \ \delta (\Delta s[\tilde{\bf \Gamma},\tilde{\bf C}]=-a \ k_{\text{B}}) \ {\cal D}[{\bf \Gamma}] \ . \\
\end{eqnarray*}
In the second and last equality we exploited Eq.~(\ref{sDef}).
Using the trivial fact that the path integration can be reordered
in time,
we arrive at
\begin{equation*}
 \text {Prob}(\Delta s=a \ k_{\text{B}}) = \Exp{a} \ \int P[{\bf \Gamma},\tilde {\bf C}]
                                                          \ \delta (\Delta s[{\bf \Gamma},\tilde{\bf C}]=-a \ k_{\text{B}}) \ {\cal D}[{\bf
                                                          \Gamma}] \ . \label{EntProdRes1}
\end{equation*}
Comparing this result with the probability
\begin{equation*}
\text {Prob}(\Delta s=-a \ k_{\text{B}}) = \int P[{\bf
\Gamma},{\bf C}] \ \delta (\Delta s[{\bf \Gamma},{\bf C}]=-a \
k_{\text{B}}) \ {\cal D}[{\bf
                         \Gamma}] \label{EntProdRes2}
\end{equation*}
to observe a reduction $\Delta s = -a \ k_{\text{B}}$, yields the
detailed FT (\ref{FT1}) for any symmetric protocol, ${\bf C} =
\tilde {\bf C}$. Therefore the detailed FT holds not only in the
steady state, which the system reaches under time-independent
forcing, ${\bf C}(t)=\text{const}$, but also for example in
periodically changing conditions that are symmetric with respect
to the
observed time-frame $(-\tau,+\tau)$. \\

We conclude this general derivation of FTs with the remark that
the presented formulation gives a unifying perspective on the
distinct FTs of \cite{Udo} and \cite{Maes3}. Equation~(6) in
\cite{Maes3} is generalized by Eq.~(\ref{sDef}), while the
Eqs.~(\ref{decompSys}) and (\ref{decompEnv}) correspond to
the decomposition of entropy according to the Eqs.~(5) and (14) in
\cite{Udo} respectively.

\section{Relativistic Brownian motion} \label{RBM}
The derivation of the FTs (\ref{FT1}) and (\ref{FT2}) in
Sec.~\ref{Herleitung} uses the abstract expression
(\ref{decompEnvMark}) for the entropy production $\dot
s_{\text{e}}$ in the embedding heat bath. As emphasized before,
this expression has to be evaluated for a physical process to
allow for a physical interpretation as entropy. An instructive
process is relativistic Brownian motion.
\\

To minimize technicalities, we consider first the one-dimensional
special-relativistic motion of a particle with rest mass $m$ in a
heat bath at temperature $T$. The generalization to higher spatial
dimensions is straightforward. Even if we would allow the particle
to equilibrate with its environment, the mean squared velocity may
not obey the non-relativistic law $\left<v^2 \right>=k_{\text{B}}
T/m$ in the high temperature limit, since the finite speed of
light defines an insurmountable upper bound. The
special-relativistic nonequilibrium Brownian motion, giving rise
to bounded velocity distributions, has been set forth in
\cite{Debbasch1, DunkelHaenggi1, DunkelHaenggi2} using both, the
language of stochastic differential equations (relativistic
Langevin equations) and the language of probability densities
(relativistic Fokker-Planck equations). Simulations of this
relativistic stochastic process have been applied to analyze
scattering experiments of quark-gluon plasma \cite{Hees}. As in
the familiar non-relativistic case \cite{Risken}, a deterministic
force $F_{\text{d}}$ acts on the particle in the rest frame of the
heat bath,
\begin{eqnarray}
\D p_{\text{d}} = F_{\text{d}} \ \D t =-\nu p \label{Fdet} \ \D t
\ ,
\end{eqnarray}
so that the time scale of dissipation is $1/\nu$. In the
relativistic generalization (\ref{Fdet}), the non-relativistic
momentum $mv$ is replaced by $p=p^1={mv}/{\sqrt{1-{v^2}/{c^2}}}$,
which is the spatial component of the relativistic momentum vector
$p^\alpha$. As common, Greek indices refer to temporal
($\alpha=0$) and spatial components. The signature of the
Minkowski metric tensor is $\eta_{\alpha \beta}=\eta^{\alpha
\beta}=\text{diag}(-1,1)$. Moreover, Einstein's summation
convention is invoked throughout. Since the rest mass is not
altered in elastic collisions, $p^\alpha
p_\alpha=\nolinebreak-(mc)^2=\nolinebreak\text{const}$, the change
in the momentum vector $\D p^\alpha$ is always ``orthogonal'' to
$p_\alpha$ in the sense of
\begin{eqnarray}
p_\alpha \D p^\alpha=0 \ . \label{orth}
\end{eqnarray}
This means that the classical particle cannot leave its mass shell
$p^\alpha p_\alpha=-(mc)^2$, which is nothing but its dispersion relation,
\begin{eqnarray}
E=p^0 c=\sqrt{(mc^2)^2+(pc)^2} \ . \label{DispRel}
\end{eqnarray}
The general solution of (\ref{orth}) is the projection $ \D
p^\alpha= (\delta^\alpha_\beta + \nolinebreak {p^\alpha
p_\beta}/{(mc)^2}) \xi^\beta $ of an arbitrary Lorentz vector
$\xi^\beta$. It is readily confirmed that the choice
\begin{eqnarray}
\D p_{\text{d}}^\alpha= -m \nu \left(\delta^\alpha_\beta +
\frac{p^\alpha p_\beta}{(mc)^2}\right) v_{\text{bath}}^\beta \D
\tau \label{ViscTens}
\end{eqnarray}
reduces to Eq.~(\ref{Fdet}) in the rest frame of the bath with the
bath velocity vector $v_{\text{bath}}^\alpha=(c,0)$ and the
particle's proper time $\tau$. Hence, Eq.~(\ref{ViscTens}) is the
generalized Lorentz-invariant deterministic part of the Brownian
motion\footnote{Equation~(16) in \cite{DunkelHaenggi1} contains an
identically vanishing term.}.

The description of relativistic Brownian motion is completed by
Lorentz-invariant stochastic changes $\D p^\alpha_{\text{s}}$ of
the momentum caused by the impacts of the surrounding heat bath at
temperature $T$. The derivation is guided by two principles:
first, the relativistic momentum is the proper quantity performing
a Wiener process, since it is physically exchanged and additive,
whereas the velocity is well-known not to be additive in special
relativity. The second postulate demands that the distribution is
Gaussian in the instantaneous rest frame of the particle. This
connects the relativistic Brownian motion to the non-relativistic
case. These principles determine the exchanged momenta $\D
p^\alpha_{\text{s}}$ to be distributed according to (cf. Eq.~(35c)
in \cite{DunkelHaenggi1})
\begin{eqnarray}
P_{\text{coll}}(p^\mu,\D p^\nu_{\text{s}}) = \frac{mc \
\delta\left(p_\beta \D p^\beta_{\text{s}}\right)}{2\sqrt{\pi
{\mathscr D} \D \tau}} \exp {\Big(}-\frac{\D p^\alpha_{\text{s}}\D
p_{\text{s} \; \alpha}}{4 {\mathscr D} \D \tau} {\Big)}.
\label{1u1Verteilung}
\end{eqnarray}
The Dirac distribution $\delta(p_\beta \D p^\beta_{\text{s}})$ in
(\ref{1u1Verteilung}) guarantees that the mass-shell condition
(\ref{orth}) is also fulfilled by the stochastic impacts, since
they are elastic. While the relativistic momentum $p$ is additive
and unbounded, the velocity is restricted to the open interval
$(-c,+c)$. This can be seen by the elegant relation $v/c^2=p/E$ in
the rest frame of the bath, which is equivalent to
\begin{eqnarray}
\D x = \frac{pc}{\sqrt{(mc)^2+p^2}} \ \D t \ . \label{Ortsanteil}
\end{eqnarray}

As mentioned before in the context of the general Kramers-Moyal
expansion (\ref{Strom}), the bath temperature $T$ is defined by
the Einstein relation,
\begin{eqnarray}
{\mathscr D} =\nolinebreak k_{\text{B}}T m \nu \ ,
\label{EinsteinRelation}
\end{eqnarray}
with the momentum diffusion constant ${\mathscr D}$ (cf. Eq.~(59)
in \cite{DunkelHaenggi1}).

\section{Relativistic fluctuation theorem} \label{SRResults} We have now the
manifestly Lorentz-invariant Langevin equation
\begin{eqnarray}
\D p^\alpha = \D p_{\text{d}}^\alpha + \D p_{\text{s}}^\alpha
\label{LILE}
\end{eqnarray}
with the deterministic part given by (\ref{ViscTens}) and the
stochastic part described by (\ref{1u1Verteilung}) at hand.
Specifying (\ref{LILE}) to the rest frame of the bath yields
\begin{eqnarray}
\D p = - \nu p \ \D t + \D p_{\text{s}} \label{BathLE} \ .
\end{eqnarray}
The probability density of the exchanged momenta $\D p_{\text{s}}$
is found by integrating out the $\D p^0_{\text{s}}$-component in
(\ref{1u1Verteilung}), cf.~\cite{DunkelHaenggi1}:
\begin{eqnarray}
P_{\text{coll}}(p,\D p_{\text{s}}) = \frac{\exp{\left(-\D
p_{\text{s}}^2 / (4 {\mathscr D} \sqrt{1+\frac{p^2}{(mc)^2}} \D t
) \right)}}{2\sqrt{\pi {\mathscr D} \D
t}\sqrt[4]{1+\frac{p^2}{(mc)^2}}} \ . \label{1Verteilung}
\end{eqnarray}
This exhibits the discretization dilemma: A discretization rule
has to be imposed on (\ref{1Verteilung}) since relativistic
invariance does not determine whether $p$ in (\ref{1Verteilung})
refers to the particle momentum $p_{-}$ before the collision
(pre-point rule of It\^o), to the post-point $p_{+}=p_{-}+\D p$
(H\"anggi-Klimontovich), or to the midpoint $(p_{-}+p_{+})/2$
(Fisk-Stratonovich).

The Eqs.~(\ref{Ortsanteil}), (\ref{BathLE}) and
(\ref{1Verteilung}) establish the relativistic stochastic motion
of the Brownian particle in phase space. The corresponding
transition probability is uniquely determined by the
discretization rule:
\begin{eqnarray}
P_{\text{trans}}^{\D t}\binomial{x\mapsto x+\D x}{p \mapsto p+\D
p} = \frac{\delta {\Big(}\D x - \frac{pc^2}{E}\D t
{\Big)}}{2\sqrt{\pi
{\mathscr D} E \D t / m c^2}} \nonumber \\
\times \exp{{\bigg(}-\frac{ \left( \D p + \nu p \D t - (1-\kappa)
\frac{{\mathscr D}}{mc^2} \frac{\D E}{\D p} \D t \right)^2}{4
{\mathscr D}E \D t / {mc^2}}\;{\bigg)}} \ . \label{PSVerteilung}
\end{eqnarray}
The discretization is contained in the parameter $\kappa$, $0 \le
\kappa \le 1$. H\"anggi-Klimontovich, Fisk-Stratonovich, or It\^o
correspond to the values $\kappa=\nolinebreak 0,$ $\frac{1}{2}$,
or $1$ respectively.

Let us now investigate the consequences for entropy production
arising out of the special-relativistic discretization dilemma. As
derived in Sec.~\ref{Herleitung}, the total entropy is a sum of
the particle intrinsic entropy $s_{\text{s}}=\nolinebreak
-k_{\text{B}} \ln P$ with the particle's nonequilibrium phase
space density $P({x},{p},t)$, and the external entropy
$s_{\text{e}}$ of the ambient heat bath at temperature $T$.

Inserting the probability current (\ref{Strom}) in momentum space
\begin{eqnarray}
j_p(x,p,t) = -\left(\nu p + \kappa \frac{\mathscr D}{mc^2}
\frac{\D E }{\D p} \right) P(x,p,t) -\frac{{\mathscr D} E}{m c^2} \frac{\pd P(x,p,t)}{\pd p}
\label{ExplStrom}
\end{eqnarray}
in the differential $\D s_{\text{s}}$ of the particle entropy
$s_{\text{s}}(t)=\nolinebreak -k_{\text{B}} \ln
P({x}(t),{p}(t),t)$ we find the equation of motion (generalizing
Eq.~(7) in \cite{Udo}) for $s_{\text{s}}$,
\begin{eqnarray}
\D s_{\text{s}}=\D s_{\text{s}}\vert_{\kappa=0} + \kappa \
k_{\text{B}} \ \D \ln E \ . \label{dsp}
\end{eqnarray}
Here we have isolated the second term which depends on the
discretization rule applied.

The entropy production $\D s_{\text{e}}$ in the bath follows by
contrasting the transition probabilities of the trajectory ${\bf
\Gamma}=({x},{p})$ with its time-reverse $\tilde{\bf
\Gamma}=(\tilde {x},-\tilde{p})$ to extract the irreversible part,
$\ln P^{\D t}_{\text{trans}}\left({\bf \Gamma}^- \mapsto {\bf
\Gamma}^+ \right)-\nolinebreak \ln P^{\D t}_{\text{trans}}(\tilde
{\bf \Gamma}^- \mapsto \tilde {\bf \Gamma}^+)$, causing the
dissipation (\ref{decompEnvMark}). From a brief computation
we find:
\begin{eqnarray}
\D s_{\text{e}} &=& k_{\text{B}} \ln \frac{P^{\D
t}_{\text{trans}}\binomial{x\mapsto x+\D x}{p \mapsto p+\D
p}}{P^{\D t}_{\text{trans}}\binomial{x+\D x \mapsto x}{-p-\D p
\mapsto -p}} \nonumber \\ &=& -\frac{\D E}{T} - \kappa \
k_{\text{B}} \ \D \ln E \ . \label{dsm}
\end{eqnarray}

The Eqs.~(\ref{dsp}) and (\ref{dsm}) reveal that although the
relativistic Brownian motion is physically inequivalent depending
on $\kappa$, the fluctuations of the total entropy
$s=s_{\text{s}}+s_{\text{e}}$ are independent of $\kappa$.
Explicitly, the change of the total entropy is
\begin{eqnarray*}
\frac{\D s}{k_{\text{B}}} = -\frac{\pd \ln P}{\pd t} \D t
-\frac{\pd \ln P}{\pd x} \D x + \frac{mc^2j_p}{{\mathscr D} E P}
\D p \ .
\end{eqnarray*}

Two technical comments are here in order. First, when computing
$\D s_{\text{e}}$ in (\ref{dsm})
the notation has to carefully distinguish between initial ${\bf
\Gamma}^-$ and finale state ${\bf \Gamma}^+$, and one should
consider the quotient $P({\bf \Gamma}^- \rightarrow {\bf \Gamma}^-
+ \D {\bf \Gamma}) / P({\bf \Gamma}^- + \D {\bf \Gamma}
\rightarrow {\bf \Gamma}^-)$ as done in (\ref{dsm}). Writing the
back transition in the numerator in the form $P({\bf \Gamma}^+
\rightarrow {\bf \Gamma}^+ - \D {\bf \Gamma})$ would be correct
yet unfavorable for evaluation, because common ${\bf
\Gamma}^-$-factors could not be cancelled out. In transforming
$P({\bf \Gamma}^+ \rightarrow {\bf \Gamma}^+ - \D {\bf \Gamma})$
to $P({\bf \Gamma}^- + \D {\bf \Gamma} \rightarrow {\bf
\Gamma}^-)$ the known spurious drift of the multiplicative
coupling has to be taken into account \cite{Risken}. Second, the
discretization term in (\ref{dsm}) can be absorbed by defining a
more complicated fluctuation-dissipation theorem, however in this
article we use exclusively the Einstein relation
(\ref{EinsteinRelation}).

The path integration of the results (\ref{dsp}) and (\ref{dsm})
according to Sec.~\ref{Herleitung} yields the detailed FT
(\ref{FT1}) for time-symmetric environments, and the integral FT
(\ref{FT2}) for arbitrary environmental conditions, with entropy
fluctuations $\Delta s$ observed over finite time. Therewith we
have proven relativistic FTs that are unaffected by the
discretization dilemma.

Furthermore, we are now in a position to address the physical
choice of $\kappa$ by virtue of the FT. Because of energy
conservation, the energy $-\D E$ in (\ref{dsm}) lost by the
particle equals the heat $\D Q$ gained by the ambient bath:
\begin{eqnarray}
\D s_{\text{e}} = \frac{\D Q}{T} - \kappa \ k_{\text{B}} \ \D \ln
E \ . \label{dsm2}
\end{eqnarray}
In the non-relativistic regime the particle energy $E=\nolinebreak
mc^2+E_{\text{kin}}$ is dominated by the energy of the rest mass
$m$ so that the second term in (\ref{dsm2}) vanishes for $mc^2\gg
E_{\text{kin}}$,
\begin{eqnarray*}
  \D \ln E=\frac{E_{\text{kin}}}{mc^2+E_{\text{kin}}} \ \D \ln E_{\text{kin}} \
  ,
\end{eqnarray*}
and we recover the non-relativistic FTs \cite{Udo}. At arbitrary
relativistic energies (\ref{DispRel}) the H\"anggi-Klimontovich
rule, $\kappa=0$, entails the correct expression for the entropy
\begin{eqnarray}
\D s_{\text{e}} =  \frac{\D Q}{T} \ , \label{result1}
\end{eqnarray}
which is produced in the heat bath.

\subsection{Generalizations in the framework of special
relativity} To generalize the FTs to $n$ spatial dimensions,
momentum and force in Eq.~(\ref{Fdet}) are simply substituted by
their spatial vectors and the Greek indices in the
Lorentz-invariant Eqs.~(\ref{ViscTens}) and (\ref{1u1Verteilung})
take values up to $n$. After integrating out the temporal
component $p^0$, the distribution (\ref{1Verteilung}) is found to
contain a quadratic form $\bf A$ instead of the square in the
exponent (cf. Eq.~(15) in \cite{DunkelHaenggi2}) with tensor
components
\begin{eqnarray}
A_{ij}= \ \delta_{ij} - \frac{c^2}{E^2} p_i p_j \ .
\label{Amatrix}
\begin{array}{ccc}
\end{array}
\end{eqnarray}
The FTs follow using the fact that $\bf p$ is an eigenvector of
$\bf A$. No complications are caused by allowing an inhomogeneous
heat bath, where the temperature $T$ and the dissipation rate
$\nu$ vary in space. As far as the integral FT (\ref{FT2}) is
concerned, a bath temperature evolving in time is also permitted
(as part of the environmental condition ${\bf C}(t)$ in
Sec.~\ref{Herleitung}). Since the time-asymmetric part enters
(\ref{result1}), the dissipation rate $\nu$ may be an even
function of the momentum, $\nu({\bf p})=\nu(-{\bf p})$. This is of
physical relevance since $\nu$ is known not to be constant even
for most non-relativistic processes \cite{Raizer}. As mentioned in
the general derivation of Sec.~\ref{Herleitung}, an arbitrary
time-dependent external force $F_{\text{e}}(t)$ (being also part
of the environmental condition ${\bf C}(t)$ defined in the rest
frame of the bath) does not pose a problem. After adding
$F_{\text{e}}(t)$ to the deterministic force $F_{\text{d}}$ in
(\ref{Fdet}) we find the expression $\D s_{\text{e}}=\D Q /T$ with
the heat $\D Q=-\D E + F_{\text{e}}\D x$. This is the first law of
thermodynamics stated in the frame of the bath.

\subsection{The commuting Brownian particle} \label{ExplEx1}
We give (to the author's knowledge) the first example where the
non-Gaussian fluctuations of particle entropy $\Delta
s_{\text{s}}$, environmental entropy $\Delta s_{\text{e}}$, and
total entropy $\Delta s$ can be evaluated exactly. A complementary
method which allows the general numerical computation of
fluctuations by iteration will be proposed in Sec.~\ref{ExplEx2}.

\begin{figure} \begin{center}
\epsffile{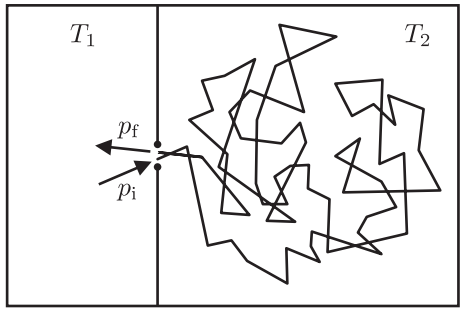} \caption{A Brownian particle commuting
between different thermal environments. The non-Gaussian
fluctuations of entropy occurring in this system can be evaluated
analytically (cf. Figs.~\ref{Graphic3} and \ref{Graphic4}).}
\label{Graphic2}
\end{center} \end{figure}

Consider two heat baths at temperatures $T_1$ and $T_2$ with a
Brownian particle moving initially in $T_1$. After the
equilibration time, its momentum $p_{\text{i}}$ (in units of $mc$)
will be distributed according to the J\"uttner-Maxwell
distribution $\varphi(p_{\text{i}},T_1)$ \cite{Juettner}, where
\begin{eqnarray}
\varphi(p,T) &=& C^{-1}  \Exp{-E(p)/T} = C^{-1}
\Exp{-\sqrt{1+p^2}/T} \nonumber
          \\ &=& \frac{\Exp{- p^2 / \left(T + T\sqrt{1+p^2}\right)}} {Z(T)}
\label{RelStat}
\end{eqnarray}
is the equilibrium solution of the Brownian motion presented in
Sec.~\ref{RBM}. The last formulation in (\ref{RelStat}) (following
from $\sqrt{1+p^2}-1=p^2/(1+\sqrt{1+p^2})$) is convenient for the
low momentum limit. To keep formulas concise, we measure heat in
units of $mc^2$ and entropy in units of $k_{\text{B}}$, so that
temperature $T$ is measured in units of $mc^2/k_{\text{B}}$. The
relativistic partition sum $Z(T)$ equals
\begin{eqnarray}
Z(T) = 2 \; \exp{(1/T)} \; K_1(1/T) \ , \label{PartSum}
\end{eqnarray}
with $K_1$ being the first modified Bessel function of the second
kind. The Brownian particle can pass to the bath $T_2$ through an
opening (cf. Fig.~\ref{Graphic2}). This opening is small enough to
keep the baths at different temperatures and to ensure that the
Brownian particle spends enough time in $T_2$ before returning to
$T_1$. So its momentum $p_{\text{f}}$ on return has become
uncorrelated to the initial value $p_{\text{i}}$ and is
distributed according to $\varphi(p_{\text{f}},T_2)$. The change
of the particle entropy $\Delta s_{\text{s}}$ and the
environmental entropy $\Delta s_{\text{e}}$ during the relaxation
of the Brownian particle in $T_2$ can be expressed using
$\varphi$:
\begin{subequations} \label{TrajEnt}
\begin{eqnarray}
\Delta s_{\text{s}} &= s_{\text{s}}(t_{\text{f}})-s_{\text{s}}(t_{\text{i}}) &= \ln \frac{\varphi(p_{\text{i}},T_1)}{\varphi(p_{\text{f}},T_2)} \label{ssEx} \\
\Delta s_{\text{e}} &= {\frac{\Delta Q}{T_2}}                                &= \ln \frac{\varphi(p_{\text{f}},T_2)}{\varphi(p_{\text{i}},T_2)} \label{seEx} \\
\Delta s_{\text{ }} &= \Delta s_{\text{s}} + \Delta s_{\text{e}} &
= \ln \frac{\varphi(p_{\text{i}},T_1)}{\varphi(p_{\text{i}},T_2)}
\label{sEx}
\end{eqnarray}
\end{subequations}
The total entropy $\Delta s$ in (\ref{sEx}) follows from the above
definitions of particle entropy
(\ref{ssEx}) and environmental entropy\footnote{Aside from the
physical expression $\Delta Q / T$ used in (\ref{seEx}), the
expression $\ln \left( {\varphi(p_{\text{f}},T)} /
{\varphi(p_{\text{i}},T)} \right)$ in terms of the equilibrium
distribution $\varphi$ is directly related to the definition
(\ref{decompEnvMark}) by the principle of detailed balance, $
\varphi(p_{\text{i}},T) \; P_{\text{trans}}^{(\Delta
t)}(p_{\text{i}} \mapsto p_{\text{f}},T) = \varphi(p_{\text{f}},T)
\; P_{\text{trans}}^{(\Delta t)}(p_{\text{f}} \mapsto
p_{\text{i}},T) $, because the baths themselves are in local
equilibrium.} (\ref{seEx}). From the resulting expression
(\ref{sEx}) we find the macroscopic Gibbs entropy,
$$\Delta S=\left<\Delta
s\right> = \int \varphi(p_{\text{i}},T_1) \ln
\frac{\varphi(p_{\text{i}},T_1)}{\varphi(p_{\text{i}},T_2)} \ \D
p_{\text{i}} \ ,$$ to equal the relative entropy of the baths,
\begin{eqnarray}
\Delta S = S_{\text{KL}}\left( T_1 \Vert T_2 \right) \ ,
\end{eqnarray}
which is also known as the Kullback-Leibler distance
\cite{Kullback}.

\begin{figure} \begin{center}
\epsffile{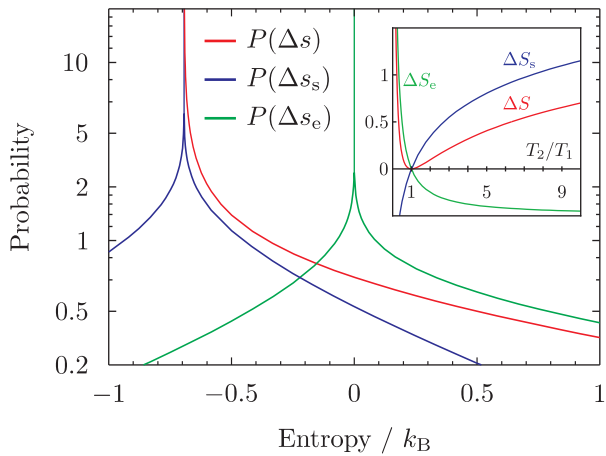} \caption{The exact expressions
(\ref{NonRelDist}) for the distribution of particle entropy
$\Delta s_{\text{s}}$ (blue), environmental entropy $\Delta
s_{\text{e}}$ (green), and total entropy $\Delta s=\Delta
s_{\text{s}}+\Delta s_{\text{e}}$ (red). The plot is for $T_1 =4
T_2 \ll mc^2/k_{\text{B}}$. The inset shows the dependence on the
temperature ratio for the macroscopic entropies
(\ref{NonRelMacEnt}): $\Delta S_{\text{s}}=\left<\Delta
s_{\text{s}}\right>$, $\Delta S_{\text{e}}=\left<\Delta
s_{\text{e}}\right>$, and $\Delta S=\Delta S_{\text{s}}+\Delta
S_{\text{e}}$ which is non-negative according to the second law of
thermodynamics (\ref{SL}).} \label{Graphic3}
\end{center} \end{figure}

The trajectory entropies (\ref{TrajEnt}) depend only on the pair
$(p_{\text{i}},p_{\text{f}})$ of initial and end point in momentum
space. Therefore the distributions $P(\Delta s)$, $P(\Delta
s_{\text{s}})$ and $P(\Delta s_{\text{e}})$ follow not from path
integrals but ordinary integrals such as $P(\Delta
s_{\text{e}})=\int \varphi(p_{\text{i}},T_1)
\varphi(p_{\text{f}},T_2) \; \delta\left(\Delta s_{\text{e}} -
\Delta s_{\text{e}}(p_{\text{i}},p_{\text{f}})\right) \; \D
p_{\text{i}}\D p_{\text{f}}$, where the expression $\Delta
s_{\text{e}}(p_{\text{i}},p_{\text{f}})$ (\ref{seEx}) is inserted
in the Dirac delta function to sum over all trajectories yielding
a certain entropy increment $\Delta s_{\text{e}}$. Because of its
physical relevance, we begin with the explicit non-relativistic
results, $T\ll mc^2/k_{\text{B}}$, when $\varphi$ (\ref{RelStat})
becomes the Maxwell-Boltzmann distribution:
\begin{subequations} \label{NonRelDist}
\begin{eqnarray}
P(\Delta s)&=&\frac{\Theta \left(A \; (\Delta
s-s_0)\right)}{\sqrt{\pi A \; (\Delta s-s_0)}} \
\Exp{-\frac{\Delta s - s_0}{A}}
\label{NonRelPs} \\
P(\Delta s_{\text{s}})&=& \frac{K_0(\vert \Delta s_{\text{s}} -
s_0 \vert)}{\pi} \label{NonRelPss} \\
P(\Delta s_{\text{e}})&=& \frac{\sqrt{\alpha}}{\pi} \; \Exp{\Delta
s_{\text{e}} \frac{1-\alpha}{2} } \; K_0 \! \left(\vert \Delta
s_{\text{e}} \vert \frac{1+\alpha}{2} \right) \label{NonRelPse}
\qquad
\end{eqnarray}
\end{subequations}
The abbreviations $A=\alpha^{-1} - 1$ and $s_0=\frac{1}{2}\ln
\alpha$ contain the dependence on the temperature ratio
$\alpha={T_2}/{T_1}$. The Heaviside step function is denoted by
$\Theta$. The distribution functions (\ref{NonRelDist}) are
plotted in Fig.~\ref{Graphic3}. With the Bessel function $K_0$
appearing in (\ref{NonRelDist}), the distributions for $\Delta
s_{\text{s}}$ and $\Delta s_{\text{e}}$ have logarithmic
divergences at $s_0$ and $0$ respectively. The distribution of
$\Delta s$ has the stronger inverse square root divergence as
$\Delta s$ approaches $s_0$ from above and vanishes below $s_0$.
From (\ref{NonRelPs}) the integral FT (\ref{FT2}) can be verified
directly, while the detailed FT (\ref{FT1}) is obviously not
fulfilled (as it has to be since the embedding temperature for the
Brownian particle changes randomly with time). We remark that
(\ref{NonRelPs}) is not simply the convolution of
(\ref{NonRelPss}) and (\ref{NonRelPse}) because $\Delta
s_{\text{s}}$ and $\Delta s_{\text{e}}$ are highly correlated.
\\
The macroscopic entropies $\Delta S_{\text{s}}=\left<\Delta
s_{\text{s}}\right>$, $\Delta S_{\text{e}}=\left<\Delta
s_{\text{e}}\right>$, and $\Delta S=\left<\Delta s\right>=\Delta
S_{\text{s}}+\Delta S_{\text{e}} \ge 0$ are the mean values of the
distributions (\ref{NonRelDist}):
\begin{subequations} \label{NonRelMacEnt}
\begin{eqnarray}
&& \Delta S_{\text{s}} = s_0 = \frac{\ln \alpha}{2} \label{NonRelSs} \\
&& \Delta S_{\text{e}} = \frac{\alpha^{-1}-1}{2} \label{NonRelSe}
\end{eqnarray}
\end{subequations}
After the Brownian particle has visited both reservoirs once, the
total macroscopic entropy increment has the symmetric form
\begin{equation}
 \Delta S(T_1\rightarrow T_2 \rightarrow T_1) = \Delta
S(T_2\rightarrow T_1 \rightarrow T_2) = \frac{(T_1-T_2)^2}{2 T_1
T_2} > 0 \ . \label{SymNonRelS}
\end{equation}

In the relativistic regime, $mc^2$ defines a third energy scale,
so that the results no longer depend only on the ratio of
temperatures. The singularity at $\Delta s=s_0$ is shifted to the
position
\begin{eqnarray}
s_0=\ln \frac{Z(T_2)}{Z(T_1)}
\end{eqnarray}
in terms of the partition sum (\ref{PartSum}). The first
relativistic correction of the partition sum is
\begin{eqnarray}
Z(T)=\sqrt{2 \pi T} \left(1 + \frac{3}{8}T + {\cal O}(T^2) \right)
\ ,
\end{eqnarray}
so that $s_0$ depends on the temperature difference $\Delta
T=T_2-T_1$ in first order:
\begin{eqnarray}
s_0=\frac{1}{2}\ln \frac{T_2}{T_1} + \frac{3}{8} \Delta T + {\cal
O}(T_1^2,T_2^2) \ .
\end{eqnarray}
In the ultra-relativistic regime, $T \gg mc^2/k_{\text{B}}$, the
partition sum becomes linear in $T$,
\begin{eqnarray}
Z(T)=2 T + 2 +  {\cal O}(1/T) \ ,
\end{eqnarray}
so that the position $s_0$ of the singularity depends on the ratio
of temperatures as in the non-relativistic limit and reaches twice
its non-relativistic value,
\begin{eqnarray}
\lim_{k_{\text{B}} T \gg mc^2} s_0=2 \lim_{k_{\text{B}} T \ll
mc^2} s_0=\ln \frac{T_2}{T_1} \ .
\end{eqnarray}
The relativistic distribution functions are sums of Bessel
functions. For example the system entropy $\Delta s$ is
distributed at arbitrary temperatures $T_1$ and $T_2$ according to
\begin{eqnarray}
P(\Delta s_{\text{s}} )&=&\frac{1}{N(T_1,T_2)} \ \left\{
\begin{array}{ll}
f(T_1,T_2,\vert \Delta s_{\text{s}}  - s_0 \vert), & \Delta s > s_0 \\
f(T_2,T_1,\vert \Delta s_{\text{s}}  - s_0 \vert), & \Delta s <
s_0
\end{array} \right. \ . \label{RelPss}
\end{eqnarray}
The normalization factor in (\ref{RelPss}) is
\begin{eqnarray}
N(T_1,T_2)=\frac{Z(T_1)Z(T_2)}{2 \sqrt{T_1 T_2} } \label{NormFac}
\end{eqnarray}
and the function $f$ in (\ref{RelPss}) is defined by the integral
\begin{eqnarray}
f(a,b,z)=\Exp{-z} \int_0^\infty \D x
\frac{\Exp{-x}}{\sqrt{x}\sqrt{x+2z}} \frac{1+ax/2}{\sqrt{1+ax/4}}
\frac{1+b (x/2+z) }{\sqrt{1+b (x/4+z/2)}} \ . \label{Complicated}
\end{eqnarray}
The non-relativistic limit (\ref{NonRelPss}) follows from
$f(0,0,z)=K_0(z)$ and $N(0,0)=\pi$. The first relativistic
corrections are
\begin{eqnarray}
\begin{array}[b]{llcl}
P(\Delta s) & = && g_0(\Delta s_{\text{s}} -
s_0) \\
& + &\overline{T} &g_1(\Delta s_{\text{s}} -
s_0) \\
& + &\Delta T &g_2(\Delta s_{\text{s}} - s_0)
\end{array} \quad & + {\cal O}(T_1^2,T_2^2,T_1 T_2) \ ,
\end{eqnarray}
with the mean temperature $\overline{T}=(T_1+T_2)/2$ and the
temperature difference $\Delta T=T_2-T_2$. We remark that the
functions
\begin{subequations} \label{HigherBessel}
\begin{eqnarray}
&&g_0(z) = \frac{K_0(\vert z \vert)}{\pi} \\
&&g_1(z) = \frac{3}{4 \pi} \left( \vert z \vert K_1(\vert z \vert)-K_0(\vert z \vert) \right) \\
&&g_2(z) = \frac{3}{8 \pi} z K_0(\vert z \vert)
\end{eqnarray}
\end{subequations}
exhibit the symmetry $T_1 \leftrightarrow T_2$ of the system,
$g_1(-z)=g_1(z), g_2(-z)=-g_2(z) $, and preserve the normalization
at any order, $\int_{-\infty}^{\infty} g_j(z) \D z =
\delta_{0,j}$.

\begin{figure} \begin{center}
\epsffile{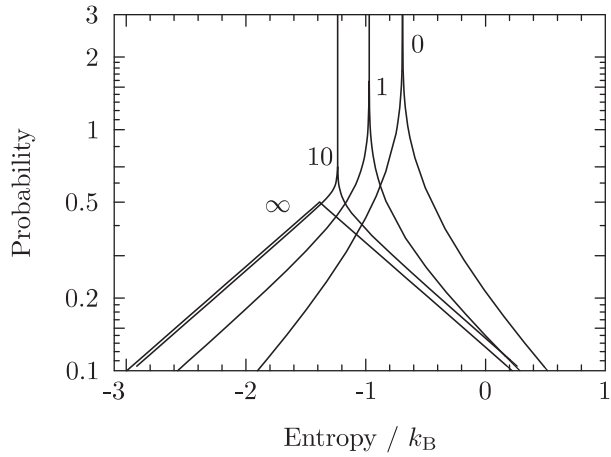} \caption{This plot shows the distribution
of system entropy, $P(\Delta s_{\text{s}})$, as we pass from the
non-relativistic regime to the ultra-relativistic regime. The
parameter attached to each graph is $k_{\text{B}} \sqrt{T_1 T_2} /
(mc^2) $ which assumes the values zero (non-relativistic limit),
$1$, $10$ and $\infty$ (ultra-relativistic limit). As we approach
the ultra-relativistic limit, the mean doubles and the spread of
fluctuations widens, but does not diverge. The logarithmic peak
reduces to a kink. Note that we are discussing the classical
relativistic regime. Quantum corrections, depending on the
particle spin, are expected when pair creation sets in.}
\label{Graphic4}
\end{center} \end{figure}

In the ultra-relativistic limit, we find $N \rightarrow 2
\sqrt{T_1 T_2}$ (\ref{NormFac}) and $f(T_1,T_2,z) \rightarrow
\sqrt{T_1 T_2} \Exp{-z} $ (\ref{Complicated}), so that
\begin{eqnarray}
P(\Delta s_{\text{s}}) = \frac{\Exp {-\vert \Delta s_{\text{s}} -
s_0\vert}}{2}
\end{eqnarray}
is an exponential distribution. It is only in the
ultra-relativistic limit that the logarithmic divergence at $s_0$
vanishes in favor of a kink (cf. Fig~\ref{Graphic4}). In the
intermediate relativistic regime ($k_{\text{B}} T \approx mc^2$)
the distribution $P(\Delta s_{\text{s}})$ has skewness. The exact
distribution (\ref{RelPss}) is shown for a fixed temperature ratio
$T_1=4T_2$ as the geometric mean $\sqrt{T_1 T_2}$ is increased
form zero (non-relativistic limit) to infinity (ultra-relativistic
limit) in Fig~\ref{Graphic4}.

\section{Generalizations in the framework of general relativity}
\label{GRResults} The monotonic increase of entropy is a
fundamental principle of physics and the universe is known to
expand, as was discovered by E.~Hubble in 1929. The discussion
whether there is a direct connection between these observations
has never stopped \cite{Hawking1, Page, Hawking2, Spec1, Spec2}.
Therefore we aspire a formulation of the FT consistent with
general relativity, but we restrict ourselves to the class of
Friedmann-Lema\^itre models, which describe a spatially homogenous
and isotropic, expanding or contracting universe. The
corresponding line element (given by the Robertson-Walker metric)
is $-\D t^2 + \D r^2$. The important difference compared to
special relativity is that the spatial part, $\D r^2$, is scaled
by a time dependent factor $R(t)$ describing the expansion or
contraction of the universe:
\begin{eqnarray}
\D r^2 = R^2(t) \ h_{ij}(\xi) \ \D \xi^i \ \D \xi^j \ .
\end{eqnarray}
The Latin indices describe spatial components numbered by $1$ to
$3$. We do not have to deal with the details of the metric tensor
$\bf h$ describing the spatial curvature. The result will be valid
for all possible geometries. The expansion rate $H(t)=\dot
R(t)/R(t)$, named Hubble function, is one of the most important
quantities in cosmology and its present value is a direct
observable \cite{Peacock}. The typical frame for a cosmic heat
bath is the frame of the cosmic microwave background.

\subsection{Cosmological fluctuation theorem} \label{GRRes1}
\begin{figure} \begin{center}
\epsffile{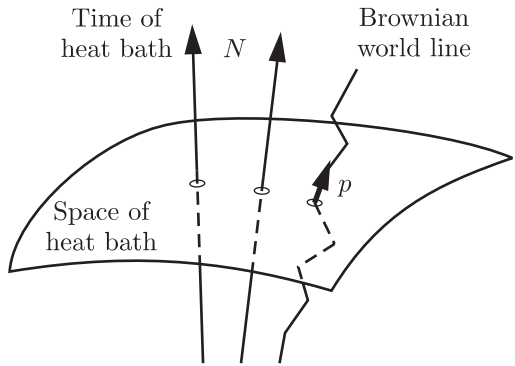} \caption{A sketch of spacetime showing a
spatial slice of the heat bath at fixed time and the world line of
a Brownian particle in a (locally) expanding universe.}
\label{Graphic1}
\end{center} \end{figure}

In general relativity, the correct equations of motion include the
covariant differential $\DD p$ of the momentum. (Denoting by $p$
the 4-vector, the components of $\DD p$ are $\DD p^\alpha=\D
p^\alpha + \Gamma_{\mu \nu}^\alpha p^\mu \D x^\nu $.) Its spatial
components replace the left hand side of (\ref{BathLE}) and can be
split up into a spatially covariant part, ${}^{(3)}\DD {\bf p}$,
and a contribution due to the time-dependent scaling:
\begin{eqnarray}
\DD {\bf p} = {}^{(3)}\DD {\bf p} + H(t) \ {\bf p} \ \D t \ . \label{RelEqMot}
\end{eqnarray}
Therefore the covariant Langevin equation, generalizing
Eq.~(\ref{BathLE}) to be valid in an expanding or contracting
universe of arbitrary spatial geometry, reads
\begin{eqnarray}
{}^{(3)} \DD {\bf p} =  - \left[\nu({\bf p},t) + H(t)\right] {\bf
p} \ \D t + {}^{(3)} \DD {\bf p}_{\text{s}} \ .
\end{eqnarray}
Herein, $H$ enters as an additional damping term, which has caused
the cooling during the expansion of our universe and is
responsible for the cosmological red shift. The distribution of
the stochastic impacts ${}^{(3)} \DD {\bf p}_{\text{s}}$ is found
after substituting  $h_{ij}$ for the Euclidean metric $\delta_{ij}$ in (\ref{Amatrix}). Applying the time-reversal map, we find that Eq.~(\ref{result1}) gains a second term due to the cosmic expansion:
\begin{eqnarray}
\D s_{\text{e}} &=& -\frac{\D E}{T} \ - \frac{\Vert{\bf p} \Vert^2}{E T} \ \D \ln R \nonumber \\
&=& -\frac{\D E}{T} - H \frac{\left({\bf p},\D {\bf r}\right)}{T} \label{result2} \\
&=& \D s_{\text{e}}^{\text{(particle)}} + \D s_{\text{e}}^{\text{(cosmic)}} \ . \nonumber
\end{eqnarray}
The numerator $\left({\bf p},\D {\bf r}\right)$ in (\ref{result2})
is the canonical line integral (canonical one-form) in phase
space. The integral FT (\ref{FT2}) extends to an expanding ($H>0$)
or contracting ($H<0$) spacetime when this second term is taken
into account. It has a clear geometric interpretation: the Hubble
function is the external curvature of space,
\begin{eqnarray}
\DD N = H \ \D {\bf r}\ , \label{DivConvergence}
\end{eqnarray}
with $N$ being the time-like normal vector to the space of the
heat bath as depicted in Fig.~\ref{Graphic1}. This permits the
second term in (\ref{result2}) to be written as
\begin{eqnarray*}
\D s_{\text{e}}^{\text{(cosmic)}} = -\frac{(p,\DD N)}{T} \ .
\end{eqnarray*}
Since the particle energy $E=p^0=-p_0=-(p,N)$ is the zero
component of the 4-vector $p$, the first term in (\ref{result2})
equals the differential
\begin{eqnarray*}
 \D s_{\text{e}}^{\text{(particle)}} =  \frac{\D (p,N)}{T} =  \frac{(\DD p,N)+(p,\DD N)}{T} \ ,
\end{eqnarray*}
such that the sum of both terms is
\begin{eqnarray}
\D s_{\text{e}} = \frac{(\DD p,N)}{T} \ . \label{result3}
\end{eqnarray}
It is natural to think of the numerator $(\DD p,N)$ as the heat
$\D Q = T \D s_{\text{e}}$ exchanged with the bath, since it is
the projection of the exchanged 4-momentum $\DD p$ on the local
energy component $N$ of the heat bath.

Cosmology is an example for the breaking of the first law, $-\D E
= \D (p,N) \neq  (\DD p,N) = \D Q$, by non-static metrics. So we
find ourselves in a remarkable situation: There is no first law in
cosmology, while the second law and furthermore the integral FT
hold.

The isolated cosmological entropy term $\D
s_{\text{e}}^{\text{(cosmic)}}$ in (\ref{result2}) would indeed
undergo a change of sign if the expansion turned into contraction.
But in the entire bath entropy (\ref{result3}) the geodesic flow
$N$ enters as a projection, which does not imply a change of sign
if $N$ was to contract. Eventually, a decreasing total entropy
$s=s_{\text{s}}+s_{\text{e}}$ is always exponentially unlikely as
expressed by the integral FT (\ref{FT2}).

\subsection{Additional theorems for the Einstein-de~Sitter universe} \label{GRRes2}
So far, we applied the time-reversal transformation to arrive at
$\left<\Exp{-\Delta s}\right>=1$. As emphasized at the outset of
the general derivation in Sec.~\ref{Herleitung}, we are free to
choose any other transformation from the mathematical point of
view. Then the function in the exponent will no longer equal the
entropy $\Delta s$. For instance, if the system is invariant under
the chosen transformation, we will get the trivial result
$\left<\Exp{0}\right>=1$. But for physically sensible
transformations, the FT will remunerate us with non-trivial
relations.
In order to derive a FT that contains the cosmic expansion rate
$H$, let us choose a local time-reversal transformation, which
acts only on the local particle dynamics and leaves the sign of
the global cosmic expansion rate $H$ unchanged. Repeating the
computation of Sec.~\ref{GRRes1} with the transformation $\tilde H
= H$ (local time reversal) instead of $\tilde H = -H$ (global time
reversal) yields
\begin{eqnarray}
\left<\Exp{-(\Delta s + \Delta h)}\right> =1, \text{ with }
 \Delta h = \frac{A H}{T}. \label{SecondGRFT}
\end{eqnarray}
The additional term $\Delta h$ is proportional to the Hubble
constant, the inverse temperature and the action $A = \int \left(
({\bf p},{\bf v}) - \Delta E\right) \ \D t $ of the energy change
$\Delta E = {\dot E}/{\nu}$.

This demonstrates that the FT is an efficient technique to design
relations that include those physical observables, which are most
interesting for a given system or experiment. The second general
relativistic FT (\ref{SecondGRFT}) holds in addition to
(\ref{FT2}). At first glance one might be surprised that there is
an infinity of FTs, all constraining the fluctuations of $\Delta
s$. But since the distribution function $P(\Delta s)$ is a point
in the infinite dimensional (Banach) space of integrable
functions, there has to be an infinity of physical constraints to
determine $P(\Delta s)$ uniquely.

In many interesting stages of the cosmic evolution, such as the
early (hypothetical) inflationary phase and the future phase of
accelerated expansion, the size of the universe grows
exponentially with time so that $H$ is constant. During these
periods the cosmic impact on the local relativistic Brownian
motion with (\ref{RelEqMot}) is time-independent. We can therefore
immediately infer from the general derivation in
Sec.~\ref{GeneralDFT}, that for these phases of the cosmic
evolution the stronger detailed formulations of the FT hold as
well.

\subsection{The expanding universe} \label{ExplEx2}

The cosmological FTs of the Secs.~\ref{GRRes1} and \ref{GRRes2}
restrict the entropy fluctuations $\Delta s$ caused by a
relativistic particle. In this section we compute the detailed
distribution of fluctuations explicitly for evolving cosmic
environments.
\\

The entropy change $\D s = \D s_{\text{s}} + \D
s_{\text{e}}^{\text{(particle)}} + \D
s_{\text{e}}^{\text{(cosmic)}}$ has contributions of the system
entropy, $s_{\text{s}}=- \ln P$, and by heat exchange, $\D
s_{\text{e}}= {\D Q}/{T} $. Since there is no first law, $\D E +
\D Q \neq 0$, for the time dependent cosmic metric, the heat
contribution $\D s_{\text{e}}$ splits up in a term due to the
change of particle energy, $\D
s_{\text{e}}^{\text{(particle)}}=-{\D E}/{T}$, and a cosmological
term, $\D s_{\text{e}}^{\text{(cosmic)}}=-{{p^2} H \D t}/{(E T)}$,
as derived in Eq.~(\ref{result2}). The method to compute
distributions of fluctuations will be presented for the particle
term, $s_{\text{e}}^{\text{(particle)}}$, which we abbreviate by
$s_{\text{p}}$. It is straight forward to apply the method to the
other terms.

To compute the distribution $P(\Delta s_{\text{p}},\Delta t)$ of
produced entropy $\Delta s_{\text{p}}$, we have to sum up $\D
s_{\text{p}}=-{\D E}/{T(t)}$ over the observation time $\Delta t$.
Therefore we have to evolve the process $p(t)$ while book keeping
the change of entropy $s_{\text{p}}$. This is done by extending
the Fokker-Planck equation to evolve the joint distribution
$P(p,s_{\text{p}},t)$. The evolution of entropy $s_{\text{p}}$ is
directly related to the dynamical variable $p$ by the differential
$\D s_{\text{p}}=-{\D E}(p)/{T(t)}$, since relativistic Brownian
motion is restricted to the mass-shell (\ref{DispRel}). The
similar evolution of $P(p,E,t)$ is easily determined. One method
is to include the Helfand moments $\left< \D E \right>$, $\left<
\D E^2 \right>$, and $\left< \D E \ \D p \right>$ into the
probability current (\ref{Strom}), from which we find the
Fokker-Planck equation $\pd_t P+ \pd_p j_p + \pd_E j_E=0$ for
$P(p,E,t)$. The correlation $\left< \D E \ \D p \right>$ is
important because $\D E$ is not independent from $\D p$ on the
mass-shell. Equivalently, we can proceed using the Fokker-Planck
equation $\pd_t P+ \pd_p j_p=0$ for $P(p,t)$ with the current
(\ref{ExplStrom}) and substitute every differentiation $\pd_p$ by
$\pd_p+\frac{\pd E}{\pd p} \pd_E$ so that the probability current
is tangential to the mass-shell. After identifying
$\pd_{s_\text{e}}=-T \pd_E$ (\ref{result1}) we arrive at the
Fokker-Planck equation
\begin{eqnarray}
\frac{1}{\nu}\pd_t P
&=& F_0(\pd_p-\frac{E'(p)}{T} \pd_{ s_{\text{p}} } ) \ P \nonumber \\
&=&\left[F_0(\pd_p) -  F_1(\pd_p) \ \pd_{s_{\text{p}}} + F_2 \
\cdot \ \pd_{s_{\text{p}}}^2\right]P
  \label{EntropyFokkerPlanck}
\end{eqnarray}
for the distribution $P(p,s_{\text{p}},t)$. The momentum
operator is
$$ F_0(\pd_p)=\lambda+(\lambda+T/E) p \pd_p + E T \pd_p^2 \ . $$
The entropic extensions of the Fokker-Planck Eq.~(\ref{EntropyFokkerPlanck}) are
$$ F_1(\pd_p)=1+\lambda F_2+2p\pd_p \text{ and } F_2=p^2/(E T) \ . $$
The function $\lambda(t)=1+H(t)/\nu$ contains the cosmic driving
by expansion. This  function of time is deterministic since we can
safely neglect the back reaction of our tiny system on the cosmic
evolution. The entropy fluctuations $P(\Delta s_{\text{p}},\Delta
t)$ follow from (\ref{EntropyFokkerPlanck}) when solved for the
initial condition
\begin{eqnarray}
  {{P(p,{s_{\text{p}}},t)}\vert}_{t=0}=\delta({s_{\text{p}}}) \ P_0(p) \label{InitialCond}
\end{eqnarray}
and after integrating out the momentum $p$:
\begin{eqnarray}
  P(\Delta {s_{\text{p}}},\Delta t)=\int_{\R} P(p,{\Delta s_{\text{p}}},\Delta t) \ \D
  p\ . \label{EntropVert}
\end{eqnarray}
The Fokker-Planck Eq.~(\ref{EntropyFokkerPlanck}) is solved by
orthogonal functions. We expand the distribution
$P(p,s_{\text{p}},t)$ in a series of Hermite polynomials with
respect to the entropy dependence, so that the two-dimensional
Fokker-Planck (\ref{EntropyFokkerPlanck}) for
$P(p,s_{\text{p}},t)$ reduces to an one-dimensional system for the
coefficients $a_k(p,t)$. The coefficients $a_k(p,t)$ are simple
linear combinations of the moments $M_l(p,t)$,
$$ M_l(p,t) = \int P({p,s_{\text{p}}},t) \ s_{\text{p}}^l \ \D s_{\text{p}} \ ,$$
so that the singular initial condition (\ref{InitialCond}) are
represented by $M_0(p,0)=P_0(p)$ and $M_l(p,0)=0$ for all $l>0$ in
a regular way. From (\ref{EntropyFokkerPlanck}) follows after
integrating by parts a hierarchy of differential equations for the
moments $M_l(p,t)$:
\begin{equation}
\frac{1}{\nu}\pd_t M_l = F_0(\pd_p) M_l + l \ F_1(\pd_p) M_{l-1} + l(l-1) \ F_2 \  M_{l-2} \ . \label{MomentFokkerPlanck}
\end{equation}
The case $l=0$ reduces to the Fokker-Planck equation for the
momentum, $M_0(p,t)\equiv P(p,t)$. Since (\ref{MomentFokkerPlanck}) is a
parabolic differential equation, numerical solutions for the
$M_l(p,t)$ can be obtained by standard techniques. Integrating $p$,
we have the moments $m_l(\Delta t)=\int M_l(p,\Delta t) \ \D p$ for
the distribution of entropy (\ref{EntropVert}). After computing
iteratively a sufficient number of moments $m_l$, the probability
distribution for the entropy (\ref{EntropVert}) can be reconstructed by the algorithm presented in Appendix~\ref{Ap}.

\begin{figure} \begin{center}
\epsffile{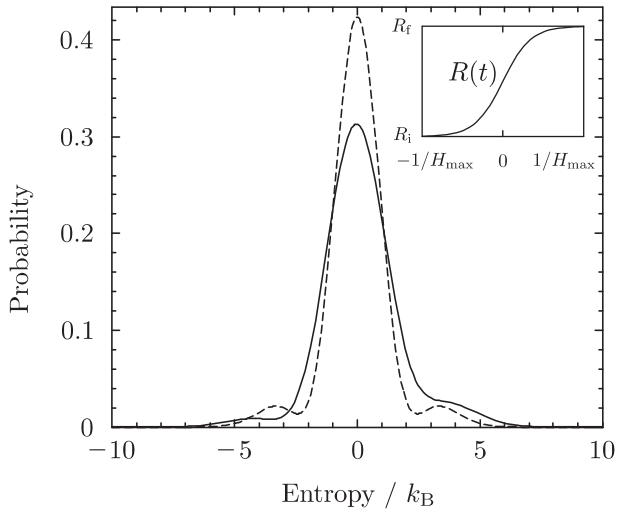} \caption{The distribution of particle
entropy $\Delta s_{\text{p}}$ is shown for a cosmic inflationary
phase. The system starts in equilibrium at time $t=-1/\nu$ with
temperature $T_{\text{i}}$, undergoes a period of inflation
centered at $t=0$ (cf. the inset), and equilibrates again until
the time $1/\nu$ with the lower bath temperature $T_{\text{f}}$.
The maximum of the Hubble function is $H_{\text{max}}=100 \nu$ for
the distribution shown in solid line. The case of a static
universe with zero mean entropy is plotted in dashed line. This
system is described completely by the ratio $H_{\text{max}}/\nu$
of the cosmic expansion rate and the thermal relaxation rate, the
temperature to mass ratio $k_{\text{B}}
T_{\text{mean}}/(mc^2)=10$, and the inflation factor
$R_{\text{f}}/R_{\text{i}}=2$.} \label{Graphic5}
\end{center} \end{figure}

Let us illustrate (\ref{EntropVert}) for a universe undergoing a
transient inflation as sketched in the inset of
Fig.~\ref{Graphic5}. Such a transition of the scale factor ranging
from $R_{\text{i}}$ to $R_{\text{f}}$ according to
$$ R(\tau)=\frac{R_{\text{i}} \ \Exp{-\tau}+R_{\text{f}} \ \Exp{\tau}}{\Exp{-\tau}+\Exp{\tau}} $$
is a common toy-model for particle creation in quantum field
theory \cite{BirrellDavies}. The peak of the Hubble function shall
be $H_{\text{max}}$, so that $\tau=t    H_{\text{max}}/I$. The
inflation factor is $I=2
(\sqrt{R_{\text{f}}}-\sqrt{R_{\text{i}}})/(\sqrt{R_{\text{f}}}+\sqrt{R_{\text{i}}})$.
Neglecting quantum effects, the thermal heat bath, which may
consist of photons or other massless particles, cools proportional
to the inverse scale factor \cite{Weinberg},
$$ T(t) = \frac{T_{\text{mean}}}{ R^{-1}_{\text{i}}+R^{-1}_{\text{f}} }  \frac{2}{R(t)} \ .$$
We choose the mean temperature $T_{\text{mean}}$ in the
relativistic regime, $k_{\text{B}} T_{\text{mean}} = 10 m c^2$.
The universe inflates by the factor $R_{\text{f}}/R_{\text{i}}=2$.
The cosmic forcing of the system depends on the ratio of the
relaxation rate $\nu$ and the expansion rate $H_{\text{max}}$. For
the nonequilibrium distribution of the particle entropy $\Delta
s_{\text{p}}$ shown as solid line in Fig.~\ref{Graphic5} the
dimensionless control parameter $H_{\text{max}}/ \nu$ equals 100.
As reference, the symmetric distribution of the relativistic
equilibrium with $H_{\text{max}}=0$ is plotted in dashed line.
When $H_{\text{max}}/ \nu$ assumes the values 1, 10 and 100, the
width $\sigma_{s_{\text{p}}}$ of the distribution $P(\Delta
s_{\text{p}})$ increases monotonically, being equal to $1.45$,
$1.48$ and $1.64$ respectively. In contrast, the mean $\Delta
S_{\text{p}}$ is not monotonic and assumes the values $0.68$,
$0.69$ and $0.15$ respectively. Five moments have been computed to
construct Fig.~{\ref{Graphic5}}.

\section{Conclusions}
Relativistic FTs have been established that remain valid for high
temperatures or low masses, $mc^2\ll k_{\text{B}} T$. The integral
FT, $\left<\Exp{-\Delta s}\right>=1$, was found to hold also in
the framework of general relativity as far as the cosmic expansion
is concerned.

With the additional FT $\left<\Exp{-\Delta s-\Delta h}\right>=1$
and the numerical example of Sec.~\ref{ExplEx2} we can answer the
question raised in the introduction: yes, the cosmic expansion has
an influence on the total entropy fluctuations $\Delta s$, and the
mean values of individual terms such as the particle contribution
$\Delta s_{\text{e}}^{\text{(particle)}}$ can undergo a change of
sign for cosmic contraction. However the relation
$\left<\Exp{-\Delta s}\right>=1$ implies the second law, $\Delta
S=\left<\Delta s\right>>0$, so that for a macroscopic system the
sign of $\Delta S$ is independent of the cosmological evolution.

On the theoretical road ahead, one may expect integral FTs to hold
for arbitrary time-dependent and inhomogeneous fields, such as
gravitational waves, when the concise expression (\ref{result3})
is applied. For the process originally introduced in
\cite{Debbasch1}, the weaker inequality (\ref{SL}) has been proven
recently \cite{Debbasch2} under general conditions.

Experimentally, the change of the environmental entropy $\Delta
s_{\text{e}} = - {\Delta E}/{T}$ can be measured by detecting
single particles after a sequence of elastic collisions, i.e.
collisions without decay or excitation of internal degrees of
freedom. Such collisions are observed for heavy quarks (for
instance the charm quark) which traverse the expanding quark-gluon
plasma created by heavy-ion collisions. Nonequilibrium
thermodynamical descriptions are common for these relativistic
media \cite{Hees}. The relativistic FT is not only subject of high
energy physics and cosmology. The special-relativistic FT can be
tested with a high-precision spectroscopy experiment by shining a
laser on an excited granulate of glass or reflecting steal beads,
so that the granulate serves as a heat bath and the photons are
the ultra-relativistic ``Brownian'' particles. The environmental
entropy $\Delta s_{\text{e}} = - {\Delta E}/{T}$ then follows from
the measurement of the frequency shift $\Delta \nu = \Delta E /
h$.

The author is grateful for the comments of M.~Brinkmann,
J.~Dunkel, P.~H\"anggi, S.~Herminghaus, C.~Jarzynski, U.~Seifert,
and V.~Zaburdaev.

\begin{appendix}


\begin{figure} \begin{center}
\epsffile{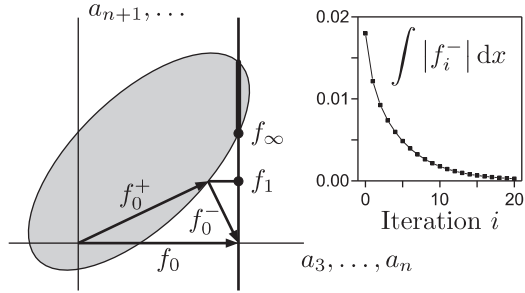} \caption{A sketch of the space
spanned by the functions (\ref{MomentSol}). The first $n$
coefficients $a_3,\dots,a_n$ are predetermine by the known moments
(vertical line). The higher coefficients $a_{n+1},\dots$ are
determined iteratively. The shaded region represents the
non-linear (convex) space of non-negative functions. The inset
shows the rapid convergence of the algorithm.} \label{GraphicA}
\end{center} \end{figure}

\section{The truncated moment problem} \label{Ap}
To reconstructing a distribution function, we are interested in an
efficient algorithm that generates uniquely out of $n \geq 2$
given moments $m_k$ a continuous and non-negative function $f$ on
the real line, so that
\begin{eqnarray}
&& \int_\R f(x) \ x^k \ \D x = m_k \text{ for } k\leq n, \text{ and} \label{MomentEq}\\
&& \left \vert \int_\R f(x) \ H_k \ \D x \right \vert =
\text{minimal} \text{ for } k>n . \qquad \label{ComplEq}
\end{eqnarray}
In (\ref{ComplEq}) the Hermite polynomials $H_k(y)=\sum_{l=0}^k
h_{k l} \; y^l$ are written in the variable $y=(x-\nolinebreak
m_1)/\sqrt{2}\sigma$ rescaled by the width
$\sigma=\sqrt{m_2-m_1^2}$. The truncated moment problem
(\ref{MomentEq}) has to be augmented by the complementary
condition (\ref{ComplEq}) for uniqueness.
Functions solving the Eqs.~(\ref{MomentEq}) are readily given by
\begin{eqnarray}
f(x) = \frac{\Exp{-y^2}}{\sqrt{2 \pi} \sigma}
\left(1+\sum_{k=3}^\infty \frac{a_k}{2^n n!} \ H_k(y) \right) \ .
\label{MomentSol}
\end{eqnarray}
By virtue of the orthogonality of the Hermite polynomials, the
first $n$ coefficients $a_k=\left<H_k\right>_f=\int f H_k \D x =
\sum_{l=0}^k h_{kl} \; m_l$ are directly determined by the known
moments $m_l$. If one was to truncate the series (\ref{MomentSol})
after the $n$'s coefficient, the resulting function $f_0$ may take
negative values. If so, we use this negative part $f_0^-=f_0 \
\Theta (-f_0)$ of the function $f_0=f_0^+ + f_0^-$ to determine
the higher coefficients to be $a_k=-\left<H_k\right>_{f_0^-}$ for
$k>n$.
This yields a new function $f_1=f_1^+ + f_1^-$ with a smaller
negative part $f_1^-$. Iteratively one approaches the desired
solution $f_\infty$ with arbitrary precision (cf.
Fig.~\ref{GraphicA}).
\end{appendix}


\begin{thebibliography} {van Beijeren et al.}
\bibitem{Zeh}
               H.D. Zeh, {\it The Physical Basis of the Direction of
               Time} (Springer, 2001).

\bibitem{WMAP3yr}
               D.N. Spergel, {\it et.al.},
               ApJS, {\bf 148}, 175 (2003);
               C.L.~Bennett,
              Nature {\bf 440}, 1126
              (2006).

\bibitem{Hawking1}
               S.W. Hawking,
               Phys. Rev. D {\bf 32}, 2489
               (1985).

\bibitem{Page}
              D.N. Page,
              Phys. Rev. D {\bf 32}, 2496
              (1985).

\bibitem{Hawking2}
               S.W. Hawking, R. Laflamme, and G.W. Lyons,
               Phys. Rev. D {\bf 47}, 5342
               (1993).

\bibitem{Spec1}
                A.E. Allahverdyan and V.G. Gurzadyan,
                J. Phys. A {\bf 35}, 7243
                (2002).

\bibitem{Spec2}
                M. Castagnino, O. Lombardi, and L. Lara,
                Found. Phys., {\bf 33}, 877 (2003).

\bibitem{BochkovKuzovlev}
                G.N. Bochkov and Yu.E. Kuzovlev,
                Physica {\bf 106A} 443 (1981) and references therein.

\bibitem{Evans}
              D.J. Evans, E.G.D. Cohen, and G.P. Morriss,
              Phys. Rev. Lett. {\bf 71}, 2401
              (1993).

\bibitem{GallavottiCohen}
              G. Gallavotti and E.G.D. Cohen,
              Phys. Rev. Lett. {\bf 74}, 2694
              (1995);
              J. Stat. Phys. {\bf 80}, 931
              (1995).

\bibitem{Kurchan} J. Kurchan, 
              J. Phys. A {\bf 31}, 3719 (1998).

\bibitem{Zon} 
              R. van Zon and E.G.D. Cohen,
              Phys. Rev. Lett. {\bf 91}, 110601 (2003).

\bibitem{Udo}
              U. Seifert, Phys. Rev. Lett. {\bf 95}, 040602 (2005).

\bibitem{Jarzynski}
              C. Jarzynski,
              Phys. Rev. Lett. {\bf 78}, 2690 (1997);
              Phys. Rev. E {\bf 56}, 5018 (1997);
              J. Stat. Phys. 98, 77 (2000).


\bibitem{Boltzmann}
        L. Boltzmann,
        Ann. Phys. {\bf 57}, 773 (1896).

\bibitem{Gibbs}
        W. Gibbs,
        Trans. Conn. Acad. {\bf 3}, 229 (1875).

\bibitem{Debbasch1}
                F. Debbasch, K. Mallick, and J.P. Rivet, J. Stat. Phys. {\bf 88}, 945 (1997).

\bibitem{DunkelHaenggi1}
        J. Dunkel and P. H\"anggi, Phys. Rev. E {\bf 71}, 016124 (2005).

\bibitem{DunkelHaenggi2}
        J. Dunkel and P. H\"anggi, Phys. Rev. E {\bf 72}, 036106 (2005).

\bibitem{Einstein1}
                A. Einstein,
                Ann. Phys. {\bf 17}, 549
                (1905).

\bibitem{Einstein2}
                A. Einstein,
                Ann. Phys. {\bf 17}, 891
                (1905).

\bibitem{EvansSearles}
         D.J.~Evans, D.J.~Searles,
         Adv. Phys. {\bf 51}, 1529 (2002).


\bibitem{Maes2} C. Maes and K. Neto\v{c}n\'{y}, 
               J. Stat. Phys. {\bf 110}, 269 
               (2003).

\bibitem{Crooks} G.E. Crooks,
                 Phys. Rev. E {\bf 60}, 2721 (1999);
                 Phys. Rev. E {\bf 61}, 2361 (2000).
                 D. Collin {\it et al.} 
                 Nature {\bf 437}, 231 (2005).

\bibitem{Risken} H. Risken, {\it The Fokker-Planck Equation} (Springer, 1996).

\bibitem{Maes3} C. Maes and M.H. van Wieren, 
                Phys. Rev. Lett. {\bf 96}, 240601
                (2006).

\bibitem{Hees}   H. van Hees, V. Greco, and R. Rapp,
                 Phys. Rev. C {\bf 73}, 034913 (2006).

\bibitem{Raizer} Y.P. Raizer, {\it Gas Discharge Physics}   (Springer, 1991).

\bibitem{Juettner}
                 F. J\"uttner, Ann. Phys. {\bf 34}, 856 (1911).

\bibitem{Kullback}
                 S. Kullback and R. A. Leibler, 
                 Ann. Math. Stat. {\bf 22}, 79, (1951);
                 S. Kullback, {\it Information Theory And Statistics}
                 (Dover Publications, 1968).

\bibitem{Peacock}
                 J.A. Peacock, {\it Cosmological Physics} (Cambridge University
                 Press, 1999).

\bibitem{BirrellDavies} N.D.~Birrell and P.C.~Davies, {\it Quantum Fields in Curved Space}, 59 (Cambridge, 1994).

\bibitem{Weinberg} S.~Weinberg, {\it Gravitation and Cosmology}, 508 (Wiley, 1972).

\bibitem{Debbasch2}
                 M. Rigotti and F. Debbasch,
                 J. Math. Phys. {\bf 46}, 103303 (2005).

\end{thebibliography}
\end{document}